\documentclass{vgtc}                          

\usepackage{mathptmx}
\usepackage{graphicx}
\usepackage{times}
\usepackage{subfig}

\onlineid{0}

\vgtccategory{Research}

\vgtcinsertpkg

\title{[POSTER] Feasibility of Corneal Imaging for Handheld Augmented Reality}

\author{Daniel Schneider\thanks{e-mail: daniel.schneider@hs-coburg.de} %
\and Jens Grubert \thanks{e-mail: jg@jensgrubert.de}} %
\affiliation{\scriptsize Coburg University}

\abstract{Smartphones are a popular device class for mobile Augmented Reality but suffer from a limited input space. Around-device interaction techniques aim at extending this input space using various sensing modalities. In this paper we present our work towards extending the input area of mobile devices using front-facing device-centered cameras that capture reflections in the cornea. As current generation mobile devices lack high resolution front-facing cameras, we study the feasibility of around-device interaction using corneal reflective imaging based on a high resolution camera. We present a workflow, a technical prototype and a feasibility evaluation.

} 

\begin{document}


\maketitle

\section{Introduction}

Handheld touch displays are a popular medium for Augmented Reality (AR), as they allow us to interact in a multitude of mobile contexts. However, shrinking device sizes, aiming at increased mobility~\cite{729537},  often sacrifice the input space of those devices. 
One option is to extend the input space of interactive displays using sensors, leading to a decoupling between input and output space 
. Various research has sparked in the area of around-device interaction,  extending the input space to near-by surfaces or to mid-air. So far, most research focused on equipping either mobiles
, the environment
or the user with additional sensors~\cite{grubert2015multifi}. However, deployment of such  hardware modifications is hard. Market size considerations discourage application developers, which limits technology acceptance in the real-world.

We envision a future in which \textit{unmodified} mobile and wearable devices that are equipped with standard front-facing cameras allow for ample movements, including the environment around and to the sides of the device, without the need for equipping the device or the environment with additional sensing hardware. This would support expressive interaction with handheld AR devices beyond their touch displays. For example, users could simply use their hands for 3D object manipulation at the side of the devices instead of being limited by 2D input of the touch screen. Furthermore, interaction between handheld  and large public displays could be enhanced \cite{grubert2014demo}. Here, content on a public display could be selected through pointing gestures outside of the handheld touch screen, leaving the touchscreen for further interaction modes, hence, simplifying the number of mode switches needed for interaction.

In our work, we investigate the feasibility of employing corneal imaging techniques with a mobile device-centred camera and evaluate its performance.  We base our work in corneal imaging \cite{nishino2006corneal}, with it's manifold applications \cite{nitschke2013corneal} and around device interaction. Specifically, only few works in around device investigated the use of unmodified devices, but suffer from limitations such a narrow input space \cite{song2014air}.  The closest work to ours is GlassHands \cite{grubert2016glasshands}, who used a build-in front-facing camera of an unmodified handheld device and some reflective glasses, like sunglasses, ski goggles or visors. 
However, the usage of dark glasses in low-light environments may not be appropriated due to perceptual and social reasons. Hence, in this work, we propose to replace parts of their original image processing pipeline by corneal imaging techniques.

\section{Concept and Implementation}\label{sec:design}
\begin{figure} [t]
	\begin{center}
		\includegraphics[width=0.7\columnwidth]{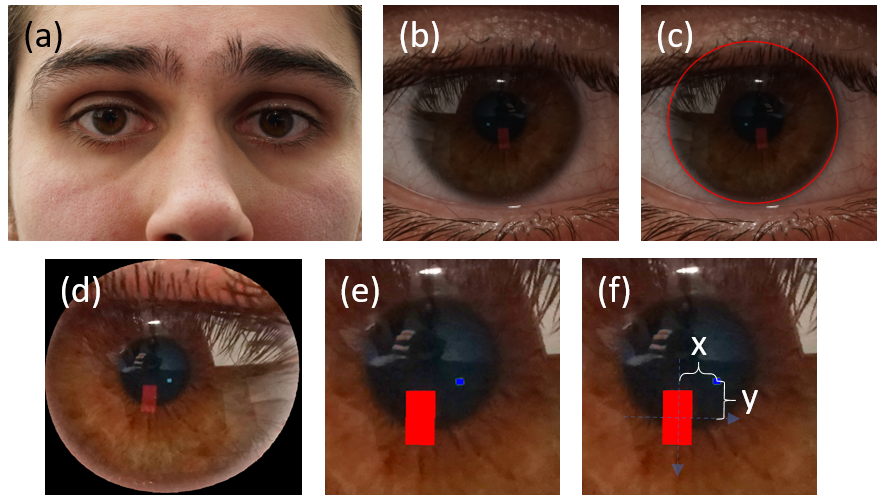}
		\caption{Results of single steps of the pipeline, (a) original camera picture, (b) eye region, (c) eye region with limbus, (d) Unwrapping result, (e) Object detection (red and blue rectangle) (f) distance of objects.}
		\label{fig:pipelineResults}
	\end{center}
\vspace{-0.5cm}
\end{figure}

To calculate the position of objects in a reflected area, one needs to be able to detect the reflection area, remove the distortion from the shape of the cornea, detect objects (in our case a handheld device and a hand) and finally to reconstruct the relative position of those objects in a common coordinate frame. 
First, an input image is searched for an eye region, then the limbus is detected and the eye pose relative to the camera is calculated. Next, the image if the corneal region is unwrapped, followed by a scene analysis step. In our case, we search for simple to detect objects and determine their relative positions in a common coordinate frame. This data is finally passed to an application which makes use of the input data. Figure \ref{fig:pipelineResults} provides an overview of the process data. The content of the pictures are related to the laboratory study in section \ref{sec:evaluation}. 
The first step of the pipeline is the detection of the eyes.  For automatic eye detection the approach of 
Kawaguchi et al. \cite{kawaguchi2005detection} can be used. The next step of the the pipeline is the limbus detection. We are using a RANSAC-based algorithm 
by Wood and Bulling \cite{wood2014eyetab} to detect the limbus, as an ellipse in the image. To calculate the pose of the eye we use an eye model of two intersecting spheres from Nitschke et al.\cite{nitschke2013corneal} for the corneal reflection the sphere of the model containing the cornea is used. One pole of the sphere is at the center of the pupil. Alternatively, more advanced eye models can be used (e.g., \cite{wood20163d}). As proposed by Nishino and Nayar \cite{nishino2006corneal}, the position and orientation of the eye can be calculated with the limbus center. The first step is to determine the distance between Pinhole of the camera and the center of the limbus.  Under weak perspective projection, the limbus in 3D space is ﬁrst orthographically projected onto a plane (average depth plane) parallel to the image plane. Since the limbus is a circle in 3D space, this average depth plane always passes through the center of the limbus. With the eye pose and the location of the eye the pixel of the reflection in the image can be projected back to the surface of the mirror (cornea). In order to process the surface of the cornea further the texture will be unwrapped. For the unwrapping we use a equirectangular projection. 
Then, scene analysis algorithms can be applied to this unwrapped image. In our case, we detect the smartphone display and the fingers in a similar manner as in GlassHands \cite{grubert2016glasshands}. The pipeline was implemented in C++ and OpenCV was used for image processing functions.

\section{Evaluation}\label{sec:evaluation}
To determine the accuracy of the position estimation of the algorithm and thus to answer the question whether it is suitable for the interaction with the mobile device, a preliminary laboratory study was carried out. Our focus was on the position estimation and not on the object recognition, hence, we opted for easily detectable colored shapes. The objects used in the study are a red rectangle (7 cm by  14 cm), which represents a mobile device and a blue square (edge length of 2 cm) as a replacement for hand or finger. During the study, the distances between the midpoints of the two objects were measured. 



We envision future built-in cameras to have high resolution imagers and, hence, did not want to limit ourselves by today's available smartphone cameras. Instead, for the recordings a Sony \(\ alpha \) 7r camera with 36 megapixel and macro lens Sony FE 90mm f / 2.8 Macro G OSS was used. 
To provide a migration path from low to high resolution imagers, the algorithm is applied to images of different sizes. For this purpose, the height and width of the images are scaled by a factor (0.5, 0.25, 0.125). The camera produces images with a resolution of 7360 x 4912 pixel. In these images the eye region is 1000 x 1000 pixel. The eye region for the factor 0.5 is 500 x 500 pixel, for 0.25 it is 250 x 250 and for the lowest it is 125 x 125, a resolution achievable with today's cameras.
The camera was placed such that the nodal point would coincide approximately with the position of a possible built-in camera (slightly above to still allow object detection).
With the study, we aimed at descriptive findings under optimal conditions and not a comparison of multiple approaches or real-world tests in mobile contexts. Hence, two expert users were invited to conduct the study under laboratory conditions. They were positioned 
with the upper body on a pedestal and viewing direction camera at a distance of 40 cm, an approximate interaction range for mobile device interaction. 
The blue square was positioned at nine different locations during the test, and an image was taken for each position. The positions are divided into three columns and three lines to the right of the red rectangle. These are parallel to the X axis (lines) and to the Y axis (columns). The columns are 10 cm, 20 cm and 30 cm from the center, and the lines are offset 10 cm up, 0 cm and 10 cm downwards. After completing the recording, the participants were thanked for their participation in the test. 
Since the limbus detection is not deterministic but merely an estimate, the algorithm is performed 20 times on each image, resulting in a total of 2 x 9 x 20 = 360 samples. 
Since the limbus detection is not deterministic but merely an estimate, the algorithm is performed 20 times on each image, resulting in a total of 2 x 9 x 20 = 360 samples. 



\subsection{Results}



The root mean square error for the full resolution image over all points is 31.13 mm (sd=20.79), for 500x500 pixel 20.69 (sd=19.76), for 250x250 pixel 28.48 (sd=35.7) and for 125x125 pixel 47.86 (sd=32.01).

\begin{figure}[!t]
	\centering
	\includegraphics[width=0.8\columnwidth]{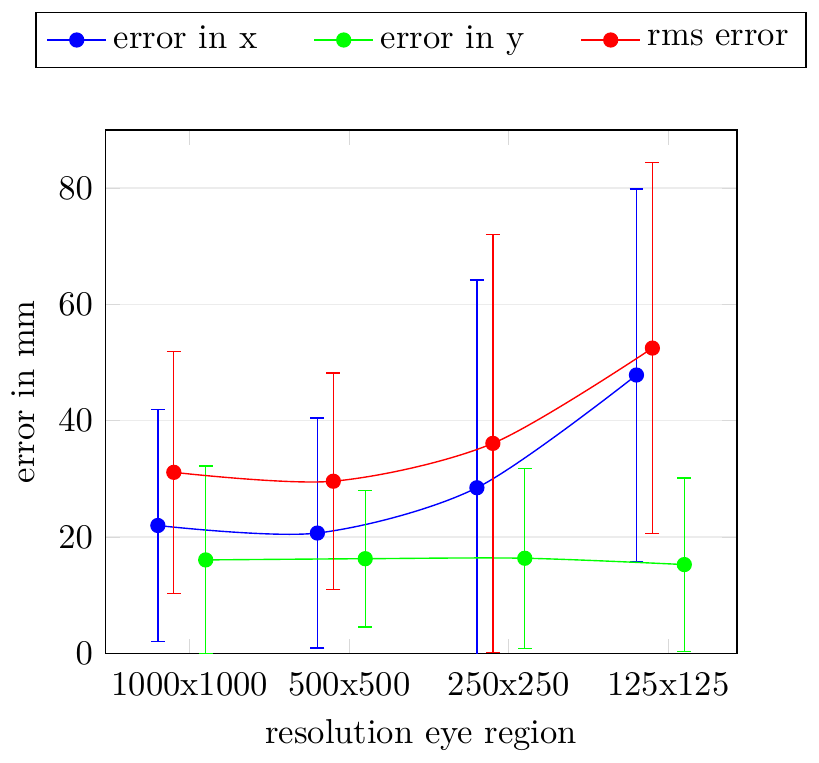}
	\caption{Errors (x, y, RMS) across all resolutions.}
	\label{fig:FehlerAll}
	\vspace{-0.5cm}
\end{figure}

The error increases with the x-coordinate (distance between the two objects in x-direction). This behavior is due to the curvature of the reflection. It is possible to explain the behavior over the resolutions (see figure \ref{fig:FehlerAll}). At the largest resolution,  a pixel corresponds to approximately 0.12°. The smaller the resolution, the larger the pixel resolutions. The smallest resolution results in a pixel corresponding to approximately 1°, which is the result of a larger deviation. In addition to the direct distance, the calculation of the projection of the red rectangle (for the construction of the plane) leads to considerable fluctuations (similar for the y-direction). 




\section{Conclusion}\label{sec:conclusion}
The evaluation has indicated that, under optimal conditions, in a region of 20 cm next to and 10 cm above or below the center point of a mobile device, positions between two sensed objects (such as a finger) could be distinguished if these are approximately 5 cm apart. To summarize, we work towards extending the input area of mobile devices using front-facing device-centered cameras that capture reflections in the cornea. To this end, we adapted corneal imaging techniques for the use in an around-device interaction pipeline. We studied the feasibility of around-device interaction using corneal reflective imaging based on a high resolution camera but also investigated the performance on lower resolution images. Our results indicate, that under optimal conditions around-device sensing could be performed with a positional resolution of ca. 5 cm. In future work, we want to extend our studies to more other settings and a larger sample size as well as optimize this approach to work on mobile phones with built-in high resolution (4k, 8k) cameras.

\bibliographystyle{abbrv}
\bibliography{template}
\end{document}